\documentclass[11pt,a4paper,twoside,groupcitations]{article}
\usepackage[T1]{fontenc}
\usepackage[ansinew]{inputenc}
\usepackage[italian,english]{babel}
\usepackage{amsfonts}
\usepackage{amsmath}
\usepackage{array}
\usepackage{amsthm}
\usepackage{amssymb}
\usepackage{graphicx}
\usepackage{subfigure}
\usepackage{braket}
\usepackage{eucal}
\usepackage{verbatim}
\usepackage[table]{xcolor}
\usepackage{caption}
\usepackage{multicol}
\usepackage{tikz}
\usetikzlibrary{arrows,shapes}
\usetikzlibrary{matrix,arrows}
\newcommand{\be}{\begin{eqnarray}}
\newcommand{\ee}{\end{eqnarray}}
\newcommand{\pro}[2]{\mbox{$\langle\, #1 \mid #2\,\rangle$}}
\newcommand{\expec}[1]{\mbox{$\langle\, #1\,\rangle$}}

\renewcommand{\d}{\mbox{${\rm d}$}} %d differenziale non corsivo in math mode
\newcommand{\lp}{\ell_{\rm p}}
\newcommand{\mpl}{m_{\rm p}}
\newcommand{\gn}{G_{\rm N}}
\newcommand{\rh}{r_{\rm H}}
\newcommand{\Rh}{R_{\rm H}}
\newcommand{\Ah}{A_{\rm H}}
\newcommand{\A}{\mathcal{A}}
\newcommand{\Th}{T_{\rm H}}
\title{\bf Thermal corpuscular black holes}
\author{Roberto~Casadio$^{ab}$\thanks{E-mail: casadio@bo.infn.it},
$\ $
Andrea~Giugno$^{ab}$\thanks{E-mail: andrea.giugno2@unibo.it},
$\ $
and
Alessio~Orlandi$^{ab}$\thanks{E-mail: orlandi@bo.infn.it}
\\
\\
{\em $^a$Dipartimento di Fisica e Astronomia, Universit\`a di Bologna}
\\
{\em via Irnerio~46, I-40126 Bologna, Italy}
\\
\\
{\em $^b$I.N.F.N., Sezione di Bologna,}
\\
{\em via B.~Pichat~6/2, I-40127 Bologna, Italy}
}
\begin{document}
\maketitle
\begin{abstract}
We study the corpuscular model of an evaporating black hole consisting of a specific quantum
state for a large number $N$ of self-confined bosons.
The single-particle spectrum contains a discrete ground state of energy $m$ (corresponding
to toy gravitons forming the black hole), and a gapless continuous spectrum (to accommodate
for the Hawking radiation with energy $\omega>m$).
Each constituent is in a superposition of the ground state and a Planckian distribution at the
expected Hawking temperature in the continuum.
We first find that, assuming the Hawking radiation is the leading effect of the internal scatterings,
the corresponding $N$-particle state can be collectively described by a single-particle wave-function
given by a superposition of a total ground state with energy $M=N\,m$ and a Planckian distribution
for $E>M$ at the same Hawking temperature. 
From this collective state, we compute the partition function and obtain an entropy which reproduces
the usual area law with a logarithmic correction precisely related with the Hawking component.
By means of the horizon wave-function for the system, we finally show the backreaction
of modes with $\omega>m$ reduces the Hawking flux.
Both corrections, to the entropy and to the Hawking flux, suggest the evaporation properly
stops for vanishing mass, if the black hole is in this particular quantum state.
%\par
%\null
%\par
%\textit{PACS - ...}
\end{abstract}
\section{Introduction}
\setcounter{equation}{0}
\label{intro}
Recently Dvali and Gomez~\cite{DvaliGomez} proposed the idea that a black hole can
be modelled as a Bose-Einstein condensate (BEC) of marginally bound, self-interacting
gravitons.
In fact, these bosons can superpose in a single small region of space, effectively giving
rise to a gravitational well, whose depth is proportional to the total number of constituents.
Furthermore, in the mean field picture, the Bogoliubov modes that become degenerate
and nearly gapless, represent the holographic quantum degrees of freedom responsible
for the black hole entropy~\cite{Bekenstein:1997bt} and the information
storage~\cite{Flassig:2012re}.
Among the merits of this model, there is thus the fact that it provides the new
perspective of describing the Hawking radiation as an emission of quanta which
are already present in the system, rather than being created out of the semiclassical
vacuum.
The result in Refs.~\cite{DvaliGomez,Flassig:2012re} have thus triggered a number
of developments~\cite{Casadio:2013hja,Kuhnel:2014oja,mueckPT,Hofmann:2014jya,Casadio:2014vja,winter}
and possible cosmological implications were also investigated in 
Refs.~\cite{Binetruy:2012kx,Kuhnel:2014gja,Dvali:2013eja,Casadio:2015xva}.
However, the lack of a geometrical description of the black hole space-time
makes it difficult to assess the presence of a horizon in this model.
\par
A possible way to make contact with the usual point-of-view of general relativity
was later proposed~\cite{Casadio:2014vja}, using the formalism
of Refs.~\cite{HWF}.
This approach can in principle be applied to any quantum state and introduces
the idea of a ``fuzzy'' gravitational radius $\Rh$ at the quantum level, starting from the
Einstein equation that determines the classical Misner-Sharp mass for spherically
symmetric systems and therefore determines the location of trapping surfaces.
For example, if a particle is in superposition of many energy eigenstates,
to each eigenstate there will correspond a different value of $\Rh$,
with a probability amplitude given by the corresponding spectral coefficient.
The particle's size could then be smaller than the mean value of $\Rh$
with finite probability.
Simple evaluations showed that only particles with a mass of the order of the
Planck scale or larger, are likely to appear in a black-hole state.
Moreover, for astrophysical masses, the relative uncertainty in the horizon size
can be acceptably small for semiclassical black holes provided the system
is made of a large number $N$ of very light condensed
bosons~\cite{Casadio:2014vja}.
In fact, in this case the horizon relative uncertainty decreases with $N$,
which supports the idea that large BECs of gravitons at the critical point
can be viewed as semiclassical black holes in the large-$N$ limit.
\par
More specifically, in Ref.~\cite{Casadio:2014vja}, we considered two
possible candidates for the quantum state representing a black hole and
its Hawking radiation: one in which the Hawking quanta have a Gaussian
distribution and one with Boltzmann distribution in energy.
Unlike the former, the latter state explicitly contains the Hawking temperature,
but cannot be used beyond perturbation theory (see Appendix~\ref{AppBoltz}). 
Hence, in this work we shall consider yet a different quantum state for the
BEC and Hawking radiation, namely we shall assume each component in
the BEC has a Planckian probability amplitude to be in an excited state.
This choice does not suffer from the drawback of the Boltzmann distribution,
and will allow us to connect nicely with the standard thermodynamic picture,
and predict the vanishing of the specific heat for small (possibly vanishing)
black hole mass.
Finally, it will also let us estimate the backreaction of the emitted quanta
by determining the horizon wave-function and mean gravitational radius 
of the system, and show that the Hawking flux vanishes at (vanishingly)
small mass, as one should expect from energy conservation.
\par
We shall start by briefly reviewing the original model of Refs.~\cite{DvaliGomez}
in the next section, and then build a specific quantum $N$-particle state
that could refine their results in section~\ref{Thermal}.
In fact, we shall show that such a state entails the Bekenstein-Hawking
entropy with a logarithmic correction and Hawking flux of outgoing particles
in section~\ref{Entropy}.
Section~\ref{HWF} is instead devoted to the calculation
of the expectation value of the gravitational radius of the system,
and the (small) deviation from the expected classical values will
be used to estimate the backreaction of the Hawking flux.
Finally, we will conclude with some comments in
section~\ref{Conclusions}.
\section{Black holes as BECs}
\setcounter{equation}{0}
\label{BEC}
Before we describe the specific quantum state we shall use to model an evaporating
black hole, let us review the basics of the model of Refs.~\cite{DvaliGomez}.
The whole construction starts from the assumption that the gravitational interaction,
approximated by the Newtonian potential
\be
V_{\rm N}(r)
\simeq
-\frac{\gn\,M}{r}
\ ,
\ee
should be strong enough to confine the gravitons themselves inside a finite volume.
If so, the gravitons will acquire an effective mass $m$ related to their characteristic
quantum-mechanical size via the Compton/de~Broglie wavelength
$\lambda_m \simeq {\hslash}/{m} = \lp\,{\mpl}/{m}$.
In fact, $N$ gravitons could superpose in a spherical volume of approximate radius
$\lambda_m$, and total energy $M=N\,m$.
Within this Newtonian approximation, the effective gravitational coupling constant
is thus given by
$\alpha =|V_{\rm N}(\lambda_m)|/N\simeq {\lp^{2}}/{\lambda_m^{2}}={m^{2}}/{\mpl^{2}}$,
and the average potential energy per graviton can be estimated as
\be
U
\simeq
m\,V_{\rm N}(\lambda_m)
\simeq
-N\,\alpha\,m
\ .
\label{eq:Udvali}
\ee
If we finally assume the gravitons are ``marginally bound'', so that
\be
E_K + U
\simeq
0
\ ,
\label{eq:energy0}
\ee
where $E_K\simeq m$ is the graviton kinetic energy, we obtain the
``maximal packing'' condition
\be
N\,\alpha
\simeq
1
\ ,
\label{eq:maxP}
\ee
which imply that the effective graviton mass and total mass of the
black hole scale like
\be
m
\simeq
\frac{\mpl}{\sqrt{N}}
\quad
{\rm and}
\quad
M
=
N\,m
\simeq
\sqrt{N}\,\mpl
\ .
\label{eq:Max}
\ee
The horizon's size, namely the Schwarzschild radius
\be
\Rh=2\,\lp\frac{M}{\mpl}
\ ,
\label{RH}
\ee
is therefore of the order of the de~Broglie length $\lambda_m\simeq \lp\,\mpl/m$ 
of the gravitons and is quantised as commonly expected~\cite{Bekenstein:1997bt},
that is
\be
\Rh
\simeq
\sqrt{N}\, \lp
\ .
\label{eq:areaquantization}
\ee
\par
For our purpose, it is of particular importance that the Hawking radiation
and its negative specific heat spontaneously result from the quantum depletion
of the condensate for the states satisfying Eq.~\eqref{eq:energy0}.
At first order, reciprocal $2 \to 2$ scatterings inside the condensate will
give rise to a depletion rate 
\be
\Gamma
\sim
\frac{1}{N^{2}}\,N^{2}\,\frac{1}{\sqrt{N}\,\lp}
\ ,
\label{eq:Grate}
\ee
where the factor $N^{-2}\sim\alpha^2$, the second factor is combinatoric
(there are about $N$ gravitons scattering with other $N - 1 \simeq N$ gravitons),
and the last factor comes from the characteristic energy of the process
$\Delta E\sim m$.
The amount of gravitons in the condensate will then decrease according
to~\cite{DvaliGomez}
\be
\dot N
\simeq
- \Gamma
\simeq
-\frac{1}{\sqrt{N}\, \lp} + {O}(N^{-1})
\ .
\label{eq:depleted-N}
\ee
As explained in Refs.~\cite{DvaliGomez}, this emission of gravitons reproduces
the purely gravitational part of the Hawking radiation and contributes to the shrinking
of the black hole according to the standard results
\be
\dot M
\simeq
\mpl\,\frac{\dot N}{\sqrt{N}}
\sim
-\frac{\mpl}{N\, \lp}
\sim
-\frac{\mpl^3}{\lp\,M^2}
\ .
\label{dotM}
\ee
From this flux one can read off the ``effective'' Hawking temperature 
\be
T_{\rm H}
\simeq
\frac{\mpl^2}{8\,\pi\,M}
\sim
m
\sim
\frac{\mpl}{\sqrt{N}}
\ ,
\label{T_H}
\ee
where the last expression is precisely the approximate value we shall
use throughout.
\section{BEC with thermal quantum hair}
\setcounter{equation}{0}
\label{Thermal}
\setcounter{equation}{0}
Like in Ref.~\cite{Casadio:2014vja}, we start by considering a system of
$N$ scalar particles, $i=1,\ldots,N$, whose dynamics is determined by a
Hamiltonian $H_i$.
We do not need to specify the latter, and will refer to such bosons as the
``toy gravitons''.
Since we want our particles to be marginally bound according to
Eq.~\eqref{eq:energy0}, we assume the single-particle
Hilbert space contains the discrete ground state $\ket{m}$, defined by
\be
\hat H_i\Ket{m}
=
m\Ket{m}
\ ,
\ee
and a gapless continuous spectrum of energy eigenstates $\Ket{\omega_i}$,
such that
\be
\hat H_i\Ket{\omega_i}
=
\omega_i\Ket{\omega_i}
\ ,
\ee
with $\omega_i>m$.
This continuous spectrum is meant to reproduce the depleted toy gravitons
that will escape the BEC.
Each particle is then assumed to be in a state given by a superposition
of $\ket{m}$ and the continuous spectrum, namely 
\be
\ket{\Psi_{\rm S}^{(i)}}
=
\frac{\ket{m}+\gamma_1\ket{\psi^{(i)}}}
{\sqrt{1+\gamma_1^2}}
\ ,
\ee
where $\gamma_1\ge 0$ is a real parameter that weights the relative probability 
amplitude for each particle to be in the continuum rather than ground state.
\par
In Ref.~\cite{Casadio:2014vja}, we considered both a Gaussian and 
a Boltzmann distribution for the continuum.
Since the Boltzmann distribution leads to inconsistent results for general values
of $\gamma_1$ (see Appendix~\ref{AppBoltz}),
we instead assume a Planckian distribution at the temperature
$T_{\rm H}=m$ [from Eq.~\eqref{T_H}] for the continuum part, namely
\be
\ket{\psi^{(i)}}
&\!\!=\!\!&
\frac{\mathcal{N}_{\rm H}}{m^{3/2}}
\int_{m}^\infty
\d \omega_i\,
{\frac{(\omega_i-m)}{\exp\left\{\frac{\omega_i-m}{m}\right\}-1}\Ket{\omega_i}}
\nonumber
\\
&\!\!\equiv\!\!&
\mathcal{N}_{\rm H}
\int_{0}^\infty
\d \mathcal{E}_i\,G(\mathcal{E}_i)\,\ket{\mathcal{E}_i}
\ ,
\label{ThermalWF}
\ee 
where the normalisation
$\mathcal{N}_{\rm H}=\sqrt{3}/\sqrt{\pi^2-6\,\zeta(3)}\simeq 1.06$,
we introduced the dimensionless variables
\be
\mathcal{E}_i
=
\frac{\omega_i-m}{m}
\  ,
\label{dimless}
\ee
and the states $\ket{\mathcal{E}_i}=m^{1/2}\,\ket{\omega_i}$, such that the identity
in the continuum can be written as
\be
\int_0^\infty
\d \mathcal{E}_i
\ket{\mathcal{E}_i}\bra{\mathcal{E}_i}
=
\int_m^\infty
{\d \omega_i}
\ket{\omega_i}\bra{\omega_i}
=
\mathbb{I}
\ .
\ee
Since we assume that this ``thermal hair'' arises because of the scatterings
among the scalars inside the BEC, the parameter $\gamma_1$ should be related
to the toy gravitons self-coupling $\alpha$, and therefore vanish for $\alpha\to 0$
or $N\to 0$.
This in turn would mean that in principle $\gamma_1$ should also depend on $N$,
for example if the maximal packing condition~\eqref{eq:maxP} remains satisfied
along the evaporation (see Ref.~\cite{winter} for a more detailed study of the scatterings
inside the BEC).
We however prefer to keep it as a free variable in this work, so that one could relate it
{\em a posteriori\/} to known features of the Hawking radiation, at least in the large
$N$ limit.
\par
In the following, we shall just assume that $\gamma_1\not=0$ yields a sufficiently
good approximation of the leading effects due to these bosons self-interactions and
therefore treat the BEC as made of otherwise free scalars.
The total wave-function of the system of $N$ such bosons will correspondingly
be approximated by the totally symmetrised product
\be
\ket{\Psi_N}
\simeq
\frac{1}{N!}\,
\sum_{\{\sigma_i\}}^N
\left[
\bigotimes_{i=1}^N
\,
\ket{\Psi_{\rm S}^{(i)}}
\right]
%=
%\frac{1}{N!}\,
%\sum_{\{\sigma_i\}}^N
%\left[
%\bigotimes_{i=1}^N
%\left(
%\ket{m}
%+
%\frac{\gamma}{\Th^{3/2}}\,\mathcal{N}_{\rm H}
%\int_{m}^\infty
%{\frac{(\omega_i-m)\, \d \omega_i}{\exp\{(\omega_i-m)/T_{\rm H}\}-1}\Ket{\psi_{\omega_i}}}
%\right)
%\right]
\ ,
\label{Nstate}
\ee
where $\sum_{\{\sigma_i\}}^N$ denotes the sum over all of the $N!$ permutations
$\{\sigma_i\}$ of the $N$ terms inside the square brackets.
Upon expanding in powers of $\gamma_1$, we obtain the exact expression
\be
\ket{\Psi_N}
&\!\!\simeq\!\!&
\frac{1}{N!}
\sum_{\{\sigma_i\}}^N
\left[
\bigotimes_{i=1}^N
\ket{m}
\right]
\nonumber
\\
&&
+
\gamma_1\,
\frac{\mathcal{N}_{\rm H}}{N!}\,
\sum_{\{\sigma_i\}}^{N}
\left[
\bigotimes_{i=2}^{N}
\ket{m}
\otimes
\int_{0}^\infty
\d \mathcal{E}_1\,
G(\mathcal{E}_1)
\ket{\mathcal{E}_1}
\right]
\nonumber
\\
&&
+
\gamma_1^2\,
\frac{\mathcal{N}_{\rm H}^2}{N!}
\sum_{\{\sigma_i\}}^{N}
\left[
\bigotimes_{i=3}^{N}
\ket{m}
\otimes
\int_{0}^\infty
\d \mathcal{E}_1\,
G(\mathcal{E}_1)
\ket{\mathcal{E}_1}
\otimes
\int_{0}^\infty
\d \mathcal{E}_2\,
G(\mathcal{E}_2)
\ket{\mathcal{E}_2}
\right]
\nonumber
\\
&&
+
\ldots
\nonumber
\\
&&
+
\gamma_1^J\,
\frac{\mathcal{N}_{\rm H}^{J}}{N!}
\sum_{\{\sigma_i\}}^{N}
\left[
\bigotimes_{i=J+1}^{N}
\ket{m}
\,
\bigotimes_{j=1}^{J}
\int_{0}^\infty
\d \mathcal{E}_j\,G(\mathcal{E}_j)
\ket{\mathcal{E}_j}
\right]
\nonumber
\\
&&
+
\ldots
\nonumber
\\
&&
+
\gamma_1^N\,
\frac{\mathcal{N}_{\rm H}^{N}}{N!}
\sum_{\{\sigma_i\}}^{N}
\left[
\bigotimes_{i=1}^{N}
\int_{0}^\infty
\d \mathcal{E}_i\,G(\mathcal{E}_i)
\ket{\mathcal{E}_i}
\right]
\ ,
\ee
where we omitted the overall normalisation constant of $1/(1+\gamma_1^2)^{N/2}$ for
the sake of simplicity.
\par
Since we are effectively including the interaction into terms proportional to powers of $\gamma_1$,
the spectral decomposition of this $N$-particle state can be obtained by defining the
total Hamiltonian simply as the sum of $N$ single-particle Hamiltonians, 
\be
\hat H
=
\bigoplus_{i=1}^N
\hat H_i
\ .
\ee
The corresponding eigenvector for the discrete ground state with $M=N\,m$ is thus
defined by
\be
\hat H\Ket{M}
=
M\Ket{M}
\ ,
\ee
and the eigenvectors for the continuum by
\be
\hat H\Ket{E}
=
E\Ket{E}
\ .
\ee
The spectral coefficients are then computed by projecting $\ket{\Psi_N}$ on 
these eigenvectors.
For $E=M=N\,m$, again neglecting an overall normalisation factor, we obtain
\be
C(M)
\simeq
\frac{1}{N!}
\bra{M}
\sum_{\{\sigma_i\}}^N
\left[
\bigotimes_{i=1}^N
\ket{m}
\right]
=
1
\ ,
\ee
as expected.
For energies above the ground state, $E>M=N\,m$, they are likewise given by
\be
C(E>M)
=
\pro{E}{\Psi_N}
\ ,
\ee
and it is again convenient to employ the dimensionless variables~\eqref{dimless},
and further introduce
\be
\mathcal{E}
=
\frac{E-M}{m}
\ ,
\ee
which lead to
\be
C(\mathcal{E}>0)
&\!\!\simeq\!\!&
\gamma_1\,
\mathcal{N}_{\rm H}\,
G(\mathcal{E})
\nonumber
\\
&&
+
\gamma_1^2\,
\mathcal{N}_{\rm H}^2
\int_0^\infty
G(\mathcal{E}_1)\,
G(\mathcal{E}-\mathcal{E}_1)\,
\d \mathcal{E}_1\,
\nonumber
\\
&&
+\ldots
\nonumber
\\
&&
+
\gamma_1^N\,
{\mathcal{N}_{\rm H}^N}
\int_0^\infty
G(\mathcal{E}_1)\, \d \mathcal{E}_1
\times\dots\times
\int_0^\infty
G(\mathcal{E}_N)\, \d \mathcal{E}_{N}
\,
\delta\!\left(\mathcal{E}-\sum_{i=1}^{N}{\mathcal{E}_i}\right)
\nonumber
\\
&\!\!\equiv\!\!&
\sum_{n=1}^N{\gamma_1^n\,C_n(\mathcal{E})}
\ ,
\label{Cseries}
\ee
where all the coefficients in this expansion can be written as
\be
C_n
=
{\mathcal{N}_{\rm H}^{n}}
\int_0^\infty
G(\mathcal{E}_1)\, \d \mathcal{E}_1
\times\dots\times
\int_0^\infty
G(\mathcal{E}_{n-1})\, \d \mathcal{E}_{n-1}\,
G\!\left(\mathcal{E}-\sum_{i=1}^{n-1}{\mathcal{E}_i}\right)
\ .
\label{BCoeff}
\ee
We then note that each integral in $\mathcal{E}_i$ is peaked around $\mathcal{E}_i=0$,
so that we can approximate
\be
G\!\left(\mathcal{E}-\sum_{i=1}^{n-1}{\mathcal{E}_i}\right)
=
\frac{\mathcal{E}-\sum_{i=1}^{n-1}{\mathcal{E}_i}}
{\exp\left\{\mathcal{E}-\sum_{i=1}^{n-1}{\mathcal{E}_i}\right\}-1}
\simeq
\frac{\mathcal{E}}
{\exp\left\{\mathcal{E}\right\}-1}
=
G(\mathcal{E})
\ ,
\ee
for $2\le n\le N$.
We thus find
\be
C_n
&\!\!\simeq\!\!&
\mathcal{N}_{\rm H}
\left(\mathcal{N}_{\rm H}\,\frac{\pi^2}{6}\right)^{n-1}
G(\mathcal{E})
=
(1.75)^{n-1}\,\mathcal{N}_{\rm H}\,
G(\mathcal{E})
\ .
\ee
Upon rescaling 
\be
\gamma\simeq 0.57\,\sum_{j=1}^N\left(1.75\,\gamma_1\right)^j
\ ,
\label{Capprox}
\ee
and switching back to dimensionful variables, we immediately obtain 
\be
C(E>M)
&\!\!\simeq\!\!&
\gamma\,
\frac{\mathcal{N}_{\rm H}}{\sqrt{m}}\,
\frac{(E-M)/m}
{\exp\left\{(E-M)/m\right\}-1}
\ .
\label{C(E)ThermApprox}
\ee
This approximation was checked numerically to work extremely well for a wide
range of $N$ (see Appendix~\ref{MC}).
\par
The result~\eqref{C(E)ThermApprox} means that we can collectively describe
the quantum state of our $N$-particle system as the (normalised) single-particle state
\be
\ket{\Psi_{\rm S}}
\simeq
\frac{\ket{M}+ \gamma\ket{\psi}}{\sqrt{1+\gamma^2}}
\ ,
\label{1PEF}
\ee
where
\be
\ket{\psi}
=
\frac{\mathcal{N}_{\rm H}}{\sqrt{m}}
\int_M^\infty
\d E\,
\frac{(E-M)/m}{\mathrm{exp}\left\{ (E-M)/m \right\}-1}
\ket{E}
\ .
\ee
In other words, from the energetic point of view, the BEC black hole effectively
looks like one particle of very large mass $M=N\,m$ in a superposition of states with
``Planckian hair''.
If we wanted to have most of the toy gravitons in the ground state,
we could just assume that $\gamma\ll 1$.
However, nothing prevents one from also considering the above
approximate wave-function for $\gamma\simeq 1$ or even larger.
\section{Partition function and entropy}
\setcounter{equation}{0}
\label{Entropy}
\par
It is now straightforward to employ the effective wave-function~\eqref{1PEF}
to compute expectations values, such as
\be
\expec{\hat H}
\!\!&=&\!\!
\frac{1}{1+\gamma^2}
\left\{
M+ \gamma^2 \,\frac{\mathcal{N}^2_{\rm H}}{m}
\int_M^\infty \left[
\frac{(E-M)/m}{\mathrm{exp}\left\{(E-M)/m\right\}-1}
\right]^2 \, E \, \d E
\right\}
\nonumber
\\
\!\!&=&\!\!
\frac{1}{1+\gamma^2}
\left[
M+ \gamma^2 \, \mathcal{N}^2_{\rm H}
\left(
M \int_0^\infty
G^2(\mathcal{E})\, \d \mathcal{E}
+m
\int_0^\infty
G^2(\mathcal{E})\, \mathcal{E} \, \d \mathcal{E}
\right)
\right]
\nonumber
\\
\!\!&=&\!\!
\frac{1}{1+\gamma^2}
\left[
M\,(1+\gamma^2) + m\,\gamma^2 \, 
\mathcal{N}^2_{\rm H}
\left(6\,\zeta(3)-\frac{\pi^4}{15}\right)
\right]
\nonumber
\\
\!\!&=&\!\!
\mpl\, \sqrt{N}
\left[
1 + \gamma^2\,\frac{\mathcal{N}^2_{\rm H}}{N}
\, 
\left(6\zeta(3)-\frac{\pi^4}{15}\right)
\right]
+ O(\gamma^4)
\ ,
\label{expE}
\ee
where we used $(1+\gamma^2)^{-1} \simeq 1-\gamma^2$ and
discarded higher powers of the parameter $\gamma$
(see Appendix~\ref{Integrals} for the integrals).
\par
Since
\be
\expec{\hat H^2}
&\!\!=\!\!&
\frac{1}{1+\gamma^2}
\left\{
M^2 + \gamma^2 \,\frac{ \mathcal{N}^2_{\rm H}}{m}
\int_M^\infty \left[
\frac{(E-M)/m}{\mathrm{exp}\left\{(E-M)/m\right\}-1}
\right]^2 \, E^2 \, \d E
\right\}
\nonumber
\\
&\!\!=\!\!&
\frac{1}{1+\gamma^2}
\left[
M^2 + \gamma^2 \, \mathcal{N}^2_{\rm H}
\left(
M^2\!\!
\int_0^\infty
\!\! 
G^2(\mathcal{E})\,
\d \mathcal{E}
+2\,M\,m\!\!
\int_0^\infty 
\!\!
G^2(\mathcal{E})\, 
\mathcal{E} \, \d \mathcal{E}
+m^2\!\!
\int_0^\infty
\!\!
G^2(\mathcal{E})\,
\mathcal{E}^2 \, \d \mathcal{E}
\right)
\right]
\nonumber
\\
&\!\!=\!\!&
\mpl^2 \, N
\left[
1 + 2 \, \frac{\gamma^2}{N} \, \mathcal{N}^2_{\rm H} 
\, \left( 6\,\zeta(3)-\frac{\pi^4}{15} \right)
+ \frac{4}{15} \, \frac{\gamma^2}{N^2} \, 
\mathcal{N}^2_{\rm H} \, \left(\pi^4-90\,\zeta(5)\right)
\right]
+O(\gamma^4)
\ ,
\ee
it is easy to obtain the uncertainty
\be
\Delta E
=
\sqrt{\expec{\hat H^2}-\expec{\hat H}^2}
&=&
\gamma\,\frac{\mpl}{\sqrt{N}} \, \mathcal{N}_{\rm H} \,
\sqrt{\frac{4}{15} \, \left(\pi^4-90\zeta(5)\right)
-\mathcal{N}^2_{\rm H} \left(6\zeta(3)-\frac{\pi^4}{15}\right)^2} 
+ O(\gamma^2)
\nonumber
\\
&\simeq&
0.76\, \gamma\,\frac{\mpl}{\sqrt{N}}
\ .
\label{uncE}
\ee
This allows us to recover the large $N$ result~\cite{Casadio:2014vja}
\be
\frac{\Delta E}{\expec{\hat H}}
\sim
\frac{\gamma}{N} + O(\gamma^2)
\ .
\ee
\par
Adopting a thermodynamical point of view, 
we can use~\eqref{expE} to estimate the partition function of the system
according to
\be
\expec{\hat H}
=
- \frac{\partial}{\partial \beta} \log{Z(\beta)}
\ ,
\ee
where $\beta=\Th^{-1}=m^{-1}\simeq {\sqrt{N}}/{\mpl}$.
We then have
\be
\expec{\hat H(\beta)}
=
\mpl^2\, \beta
\left(
1 + \frac{\A\, \gamma^2}{\mpl^2\, \beta^2}
\right)
=
\frac{\partial}{\partial \beta}
\left[
\frac{1}{2}\,\mpl^2\,\beta^2 + \A\,\gamma^2\, \log(k\,\beta)
\right]
\ ,
\ee
where $\A=\mathcal{N}^2_{\rm H}\left[6\zeta(3)-{\pi^4}/{15}\right] \simeq 0.81$,
and $k$ is an integration constant with the dimensions of a mass.
It is then straightforward to obtain
\be
\log{Z(\beta)}
=
-\frac{1}{2}\,\mpl^2\,\beta^2
-\A\,\gamma^2\, \log(\beta\,k)
\ ,
\ee
and, if we simply set $k=\mpl$, we see that
\be
Z(\beta)
=
\left(\mpl\,\beta\right)^{-\A\gamma^2}
\,
e^{-\frac{1}{2}\,\mpl^2 \beta^2}
\ ,
\ee
contains two factors.
One goes like $1/\beta=\Th$ at some power, like the thermal wave-length
of a particle does, and it is given by the contribution of the excited modes
to the energy spectrum.
This is consistent with the fact that such modes propagate freely,
since we did not associate an external potential to our model~\footnote{In
other words, we are approximating the grey-body factors for bosonic
Hawking quanta with one.}.
The exponential damping factor is instead due to the presence
of bound states localised within the classical Schwarzschild radius $\Rh$.
\par
It is now possible to compute the statistical canonical entropy
\be
S(\beta)
=
\beta^2\,
\frac{\partial F(\beta)}{\partial \beta}
\ ,
\ee
where $F(\beta)$ is the Helmoltz free energy
$F=-(1/\beta)\,\log{Z}$.
It is straightforward to get
\be
S(\beta)
=
\frac{1}{2}\,\mpl^2\,\beta^2 
- \A\,\gamma^2 \log({\mpl\,\beta})
+ \A\,\gamma^2
\ ,
\label{entrobeta}
\ee
which is nothing but the usual Bekenstein-Hawking formula~\cite{Bekenstein:1997bt}
plus a correction scaling like a logarithm.
In fact, we can write $\beta\simeq m^{-1}$ as a function of the Schwarzschild radius,
\be
\beta
=
\frac{\Rh}{2\,\lp\,\mpl}
\ ,
\ee
rescale the constant and define the area of the horizon as $\Ah=4\,\pi\,\Rh^2$,
hence yielding
\be
S
=
\frac{\Ah}{4\,\lp^2}
- \frac{\A}{2} \, \gamma^2 \,
\log\!\left({\frac{\Ah}{16\,\pi\,\lp^2}}\right)
\ .
\ee
\par
Let us conclude with a few remarks.
First off, the final expression of the entropy depends on the ``collective''
parameter $\gamma$ defined in Eq.~\eqref{Capprox}, rather than the relative probability
$\gamma_1$ for each constituent to be a Hawking mode, the latter being in turn
determined by the details of the scatterings that occur inside the BEC.
One could thus speculate that, even if we were able to detect the Hawking radiation,
the details of these interactions would hardly show directly. 
The fact remains, though, that $\gamma_1=0$ implies that also $\gamma=0$,
and and the logarithmic correction would therefore switch off if the scatterings inside
the BEC were negligible (and the Hawking radiation were absent).
Finally, we stress again that, in this corpuscular model, the total number $N$
of bosons is conserved, since both the black hole and its Hawking radiation are made
of the same kind of particles.
In connection with this, we notice that the specific heat is in our case given by
\be
C_V
=
\frac{\partial \expec{\hat H}}{\partial T}
=
-\beta^2\,
\frac{\partial \expec{\hat H}}{\partial \beta}
\simeq
-\mpl^2\,\beta^2
+\A\,\gamma^2
\ ,
\ee
which is negative for small $\gamma$ and large $\beta$ (or, equivalently, large
$N$), in agreement with the usual properties of the Hawking radiation,
but vanishes for $\beta\simeq \gamma/\mpl$, that is for $N_c\sim\gamma^2$.
If, for simplicity, we assume $1.75\,\gamma_1\sim 1$, so that $\gamma\sim N$
from Eq.~\eqref{Capprox}, we thus obtain the specific heat vanishes for
$N_c \sim 1$, as one would naively expect the Hawking radiation switches off
when there are no more particles to emit.
In fact, this result qualitatively agrees with the microcanonical picture of the
Hawking evaporation~\footnote{For the details, see Refs.~\cite{micro} and references
therein, where the black hole and its Hawking radiation where also assumed to
be of the same nature, to wit strings.}
and with the estimate of the backreaction in the next section.
\section{Horizon wave-function and backreaction}
\setcounter{equation}{0}
\label{HWF}
We can now investigate the presence of a trapping surface in our quantum state
by means of the horizon wave-function formalism.
Since the present case is very similar to the ones considered in
Ref.~\cite{Casadio:2014vja}, we refer the reader to that paper for more details
(see also Refs.~\cite{davidson} for a similar picture).
\par
The main idea behind that formalism, is to consider the relation~\eqref{RH}
that defines the gravitational radius of a spherically symmetric system as an operator
equation, that is
\be
\hat r_{\rm H}
=
2\,\lp\,\frac{\hat H}{\mpl}
\ .
\ee
Note that the classical gravitational radius $r_{\rm H}=r_{\rm H}(r)=2\,\lp\,M(r)/\mpl$
can be introduced for any spherically symmetric system with mass function
$M=M(r)$~\cite{stephani}, and it does not necessarily represent the size of a horizon.
In fact, it only represents the possible location of a trapping surface if $r_{\rm H}(r)=r$
for some value of the areal coordinate $r$.
The above definition allows one to introduce a wave-function for the
gravitational radius simply given by
\be
\Psi_{\rm H}(\rh)
\simeq
C(\mpl\,{\rh}/{2\,\lp})
\ ,
\label{HorWF}
\ee
where $C$ is precisely the spectral coefficient we computed in the previous section,
and the normalisation will be fixed assuming the Schr\"odinger scalar product
\be
\pro{\Psi_{\rm H}}{\Phi_{\rm H}}
=
4\,\pi\int_0^\infty
\Psi_{\rm H}^*(\rh)\,\Phi_{\rm H}(\rh)\,\rh^2\,\d \rh
\ .
\label{Hpro}
\ee
It is important to recall that the main result in Ref.~\cite{Casadio:2014vja} was that
the gravitational radius associated with the quantum $N$-particle states considered therein
indeed has a size very close to the expected classical value $\Rh$,
with quantum fluctuations that die out for increasing $N$, namely
\be
\expec{\hat{r}_{\rm H}}
\simeq
\Rh
+{O}(N^{-1/2})
\ ,
\ee
as one should expect  in the context of semiclassical gravity.
Since the source represented by such quantum states is by construction (mostly)
localised within $\Rh$, this shows that $r=\expec{\hat{r}_{\rm H}}$ is a horizon
for the system.
\par
Given the spectral coefficients~\eqref{C(E)ThermApprox},
the corresponding horizon wave-function reads
\be
\Ket{\Psi_{\rm H}}
=
\frac{\Ket{\Rh} + \gamma \Ket{\psi_{\rm H}}}
{\sqrt{1+\gamma^2}}
\ ,
\label{hwf}
\ee
where the state $\Ket{\Rh}$ represents the discrete part of the spectrum
and is defined so that
\be
\langle \Rh | \, \hat{r}_{\rm H} \, | \Rh \rangle
=
\Rh
\ ,
\ee
and the states
\be
\psi_{\rm H}(\rh>\Rh)
\equiv
\pro{\rh}{\psi_{\rm H}}
=
\mathcal{N}_{\rm H}\,\sqrt{\frac{N}{4\,\pi\,\Rh^3}}\,
\frac{(\rh-\Rh)/(2\,\lp\,m/\mpl)}{\exp\left\{\frac{\rh-\Rh}{2\,\lp\,m/\mpl}\right\}-1}
\ ,
\ee
which describe the contribution from the excited Hawking modes.
As usual $\Psi(\rh\leq\Rh)\simeq 0$ and the normalization is fixed by the scalar
product~\eqref{Hpro}.
The expectation value of the gravitational radius is now easily calculated, 
\be
\expec{\hat{r}_{\rm H}}
\!\!&=&\!\!
4\pi
\int_{\Rh}^\infty
\left| \Psi_{\rm H}(\rh) \right|^2 \, \rh^3 \, \d \rh
\nonumber
\\
\!\!&=&\!\!
\frac{1}{1+\gamma^2}
\left\{
\Rh +
\gamma^2\,\mathcal{N}^2_{\rm H} \, \frac{\Rh}{N^3}
\int_0^\infty
\left( \frac{\rho}{e^\rho-1} \right)^2
\, (\rho+N)^3 \, \d \rho
\right\}
\nonumber
\\
\!\!&=&\!\!
\Rh
\left[ 1 +
\frac{\gamma^2}{1+\gamma^2}\,
\frac{3\,\mathcal{N}^2_{\rm H}}{N} 
\int_0^\infty
\left(
\frac{\rho}{e^\rho-1}
\right)^2
\, \rho \, \d \rho
\right]
+ O\left(\frac{1}{N^2}\right) 
\nonumber
\\
\!\!&=&\!\!
\Rh
\left[
1 + \frac{3\,\gamma^2}{N} \, \mathcal{N}^2_{\rm H} 
\left(6\,\zeta(3)-\frac{\pi^4}{15}\right)
\right]
+ O(\gamma^4)
\ ,
\ee
where we defined
\be
\rho
\,=\,
\frac{\rh-\Rh}{2\,\lp\,m/\mpl}
\ee
together with the relation~\eqref{RH}.
We thus see that 
\be
\expec{\hat{r}_{\rm H}} - \Rh
\simeq
3\,\gamma^2 \, \frac{\Rh}{N} + O\left(\frac{1}{N^2}\right)
\ .
\ee
\par
In the same way we can compute 
\be
\expec{\hat{r}^2_{\rm H}}
=
\Rh^2
\left[
1 + 4\,\gamma^2 \, \mathcal{N}^2_{\rm H}
\left(6\zeta(3)-\frac{\pi^4}{15}\right) \frac{1}{N} 
\right]
+ O\left(\frac{1}{N^2}\right)
\ ,
\ee
which leads to
\be
\Delta\rh
=
\sqrt{|\expec{\hat{r}^2_{\rm H}}-\expec{\hat{r}_{\rm H}}^2|}
=
\Rh \, \frac{\gamma \mathcal{N}_{\rm H}}{\sqrt{N}}
\, \sqrt{2\left(6\zeta(3)-\frac{\pi^4}{15}\right)} +O\left(\frac{1}{N^2}\right)
\ee
and, omitting a factor of order one,
\be
\frac{\Delta\rh}{\expec{\hat{r}_{\rm H}}}
\sim
\frac{\gamma}{N}+O(\gamma^2)
\ ,
\ee
like we obtained in Ref.~\cite{Casadio:2014vja}.
\par
We can now note that $\expec{\hat{r}_{\rm H}}>\Rh$, albeit buy a small amount for
large $N$, which is in agreement with the fact that Hawking quanta must also contribute
to the total gravitational radius.
The constituents will therefore be bound within a sphere of radius slightly larger than the pure BEC
value $\Rh$, and the typical energy of the reciprocal $2 \to 2$ scatterings will correspondingly
be smaller.
This gives rise to a slightly smaller depletion rate, namely 
\be
\Gamma
\sim
\frac{1}{\expec{\hat{r}_{\rm H}}}
\simeq
\frac{1}{\sqrt{N}\,\lp}
\left[
1 - \frac{3\,\gamma^2}{N} \, \mathcal{N}^2_{\rm H} 
\left(6\,\zeta(3)-\frac{\pi^4}{15}\right)
\right]
\ ,
\label{eq:GrateB}
\ee
Note that, if we were to trust the above relation for small $N$,
the flux would stop for a number of quanta
\be
N_c
\simeq
{3\,\gamma^2}\, \mathcal{N}^2_{\rm H} 
\left(6\,\zeta(3)-\frac{\pi^4}{15}\right)
\simeq
2.4\,\gamma^2
\ .
\ee
If we further recall the relation~\eqref{Capprox}, and again assume for simplicity
$1.75\,\gamma_1\lesssim 1$, so that $\gamma\lesssim N$, we obtain $N_c\gtrsim 1$,
which leaves open the possibility that the flux stops for a finite number of constituents.
Of course, it is not guaranteed the above approximations still hold for such small values
of $N$, but we recall that it is a general result of the microcanonical description of the
Hawking radiation that the emitted flux vanishes for vanishingly small black hole
mass~\cite{micro}.
Furthermore, this behaviour would be in agreement with the thermodynamical analysis of
the previous section, and the vanishing of the specific heat for $N=N_c\sim 1$.
In any case, $N_c$ depends on the collective parameter $\gamma$, although the latter
is more directly related with the single-particle $\gamma_1$ for small $N$.
\section{Conclusions and outlook}
\label{Conclusions}
\setcounter{equation}{0}
We have analysed a candidate quantum state to represent an evaporating
black hole made of $N$ toy scalar gravitons, along the lines of the BEC picture
put forward in Refs.~\cite{DvaliGomez}, and improving on the cases previously
considered in Ref.~\cite{Casadio:2014vja}.
In particular, we found that a Planckian distribution at the Hawking 
temperature for the depleted modes allows one to recover the main known
features of the Hawking radiation, like the Bekenstein-Hawking entropy,
the negative specific heat and Hawking flux, for large black hole mass, or,
equivalently, large $N$.
\par
Moreover, one also obtains that the above $N$-particle state can be collectively
described very reliably by a one-particle wave-function in energy space.
From this approximate description, we obtained leading order corrections
to the energy of the system that give rise to a logarithmic contribution to the
entropy.
This contribution would ensure a vanishing specific heat for $N$ of order one,
when we expect the evaporation stops.
In fact, this qualitative result was further supported by estimating the
backreaction of the Hawking radiation onto the horizon size by means
of the horizon wave-function of the system.
Since the Hawking flux depends on the energy scale of the BEC~\cite{DvaliGomez},
which is in turn simply connected to the horizon radius, the extra contribution to the
latter by the depleted scalars precisely reduces the emission.
Upon extrapolating the picture down to small values of $N$, this 
reduction should eventually stop the Hawking radiation completely.
We think it is important to emphasise that this is exactly what one expects from the
conservation of the total energy of the black hole system~\cite{micro}.
\par
We would like to conclude by remarking that the analysis presented here
does not explicitly consider the time evolution of the system or, in other words,
it provides a possible picture of the black hole and its Hawking radiation at a given
instant in time.
One could however conjecture that, as long as the Hawking flux is relatively small,
a reliable approximation of the time evolution will be simply given by the standard 
equation~\eqref{dotM} for large mass and $N$, or by the correspondingly modified
expression that follows from Eq.~\eqref{eq:GrateB} for $N$ approaching one.
Our analysis however does not make use of any specific determination of the coefficients
$\gamma_1$ and $\gamma$ in terms of the details of the scatterings that give rise to
the Hawking radiation (and we therefore assumed no dependence of these parameters
from the number $N$).
A very interesting attempt at understanding the Hawking radiation directly from the two-
and three-body decays inside a BEC recently appeared~\cite{winter}, where
however no general relativistic effect is included.
The estimate for the flux we provide in section~\ref{HWF} could therefore be
viewed as complementary to that in Ref.~\cite{winter}, and both approaches in fact
agree in predicting a vanishing decay rate for vanishing number of constituents.
Moreover, we suspect that, for very small $N$, the BEC model could reduce to the
description of black holes as single particle states investigated in
Refs.~\cite{Casadio:2013hja,HWF} and~\cite{RN}. 
A detailed analysis of the possible transition between the two regimes is left for
future investigations.
\section*{Acknowledgments}
This work is supported in part by grant INFN-FLAG.
\appendix
\section{Boltzmann distribution}
\label{AppBoltz}
\setcounter{equation}{0}
It is easy to see that the Boltzmann distribution considered in Ref.~\cite{Casadio:2014vja}
can only be employed in the perturbative limit $\gamma_1\ll 1$.
In fact, if one replaces the Planckian distribution 
\be
G(\mathcal{E})
\to
e^{-\mathcal{E}}
\ee
in Eq.~\eqref{ThermalWF}, and expands the corresponding $C(E>M)$ in powers of $\gamma_1$
like in Eq.~\eqref{Cseries}, the analogous coefficients $C_n$ will diverge when $n\geq2$, namely
\be
C_n
\!\!&\varpropto&\!\!
\int_0^\infty \d \mathcal{E}_1 \, e^{-\mathcal{E}_1}
\int_0^\infty \d \mathcal{E}_2 \, e^{-\mathcal{E}_2}
\, \times \dots \, \times \,
\int_0^\infty \d \mathcal{E}_n e^{-\mathcal{E}_n}
\, 
\delta\left(\mathcal{E}-\sum_{i=1}^n \mathcal{E}_i\right)
\\
\notag
\!\!&=&\!\!
e^{-\mathcal{E}}
\prod_{i=1}^{n-1}{\int_0^\infty \d \mathcal{E}_i}
\to
\infty
\ .
\ee
It then follows that, for $\gamma_1\simeq 1$, and the probability for each constituent
to be a Hawking mode is not small, the Boltzmann approximation should
indeed be replaced by the Planckian distribution employed in this work. 
\section{Monte Carlo estimate of spectral coefficients}
\label{MC}
\begin{figure}
\centering
\subfigure[$C_5$]
{\includegraphics[width=8cm]{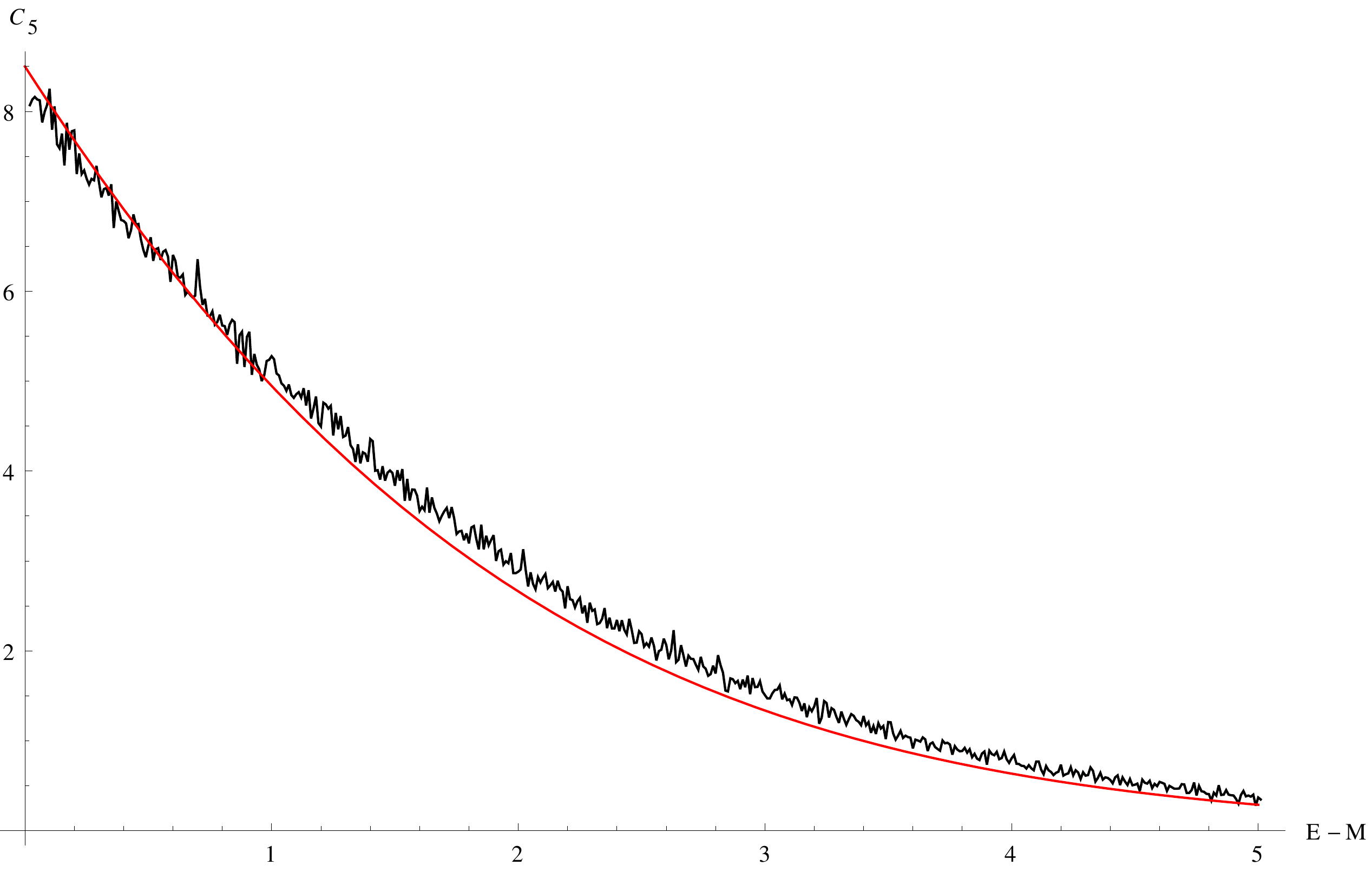}}
\hspace{1mm}
\subfigure[$C_{10}$]
{\includegraphics[width=8cm]{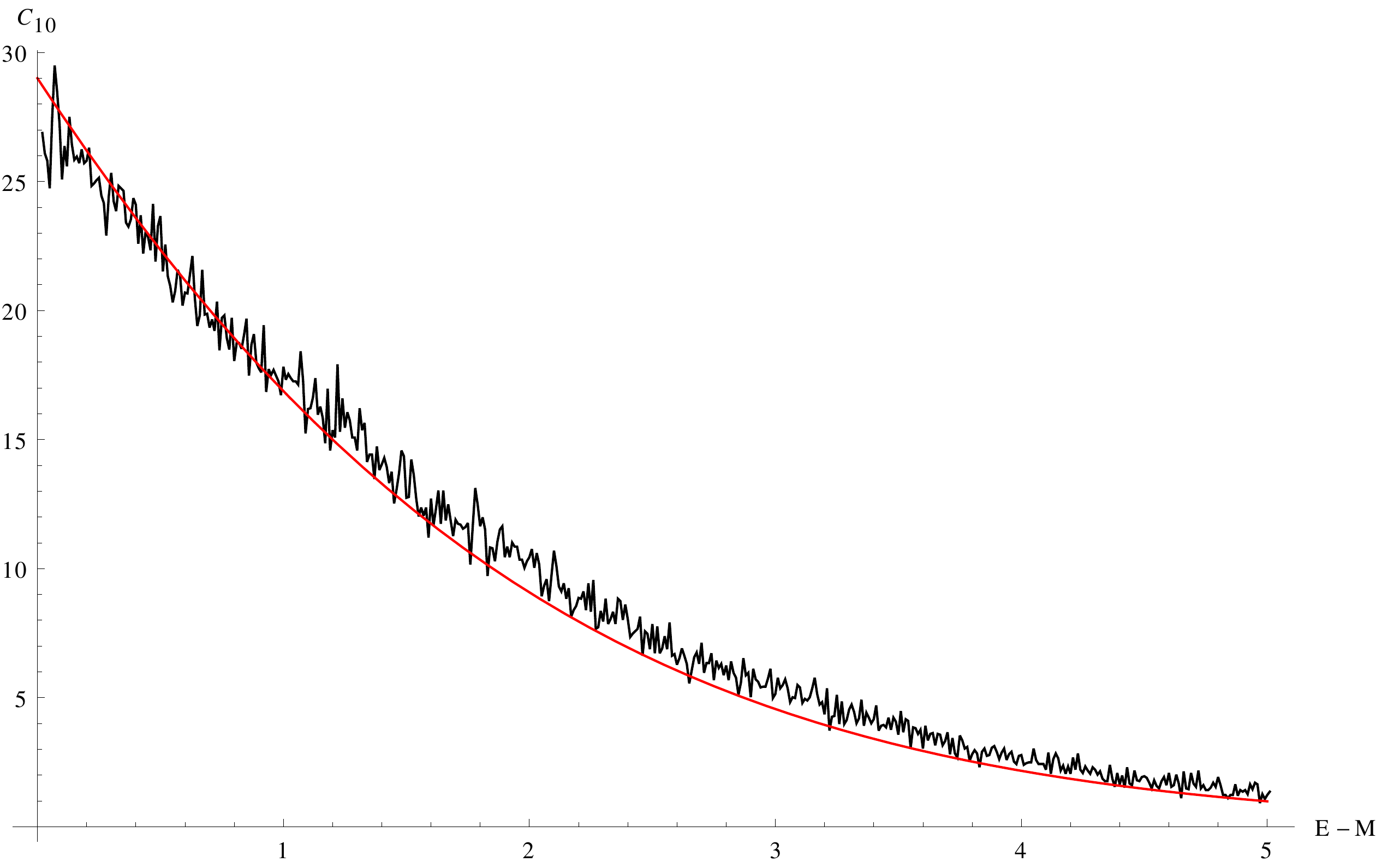}}
\caption{Numerical estimate of some coefficients in the expansion~\eqref{Cseries}
(black lines), and their analytical approximation (red lines), for $N=10$
(energies are in Planck units).}
\label{SSim}
\end{figure}

\begin{figure}[htb]
\centering
\includegraphics[width=10cm]{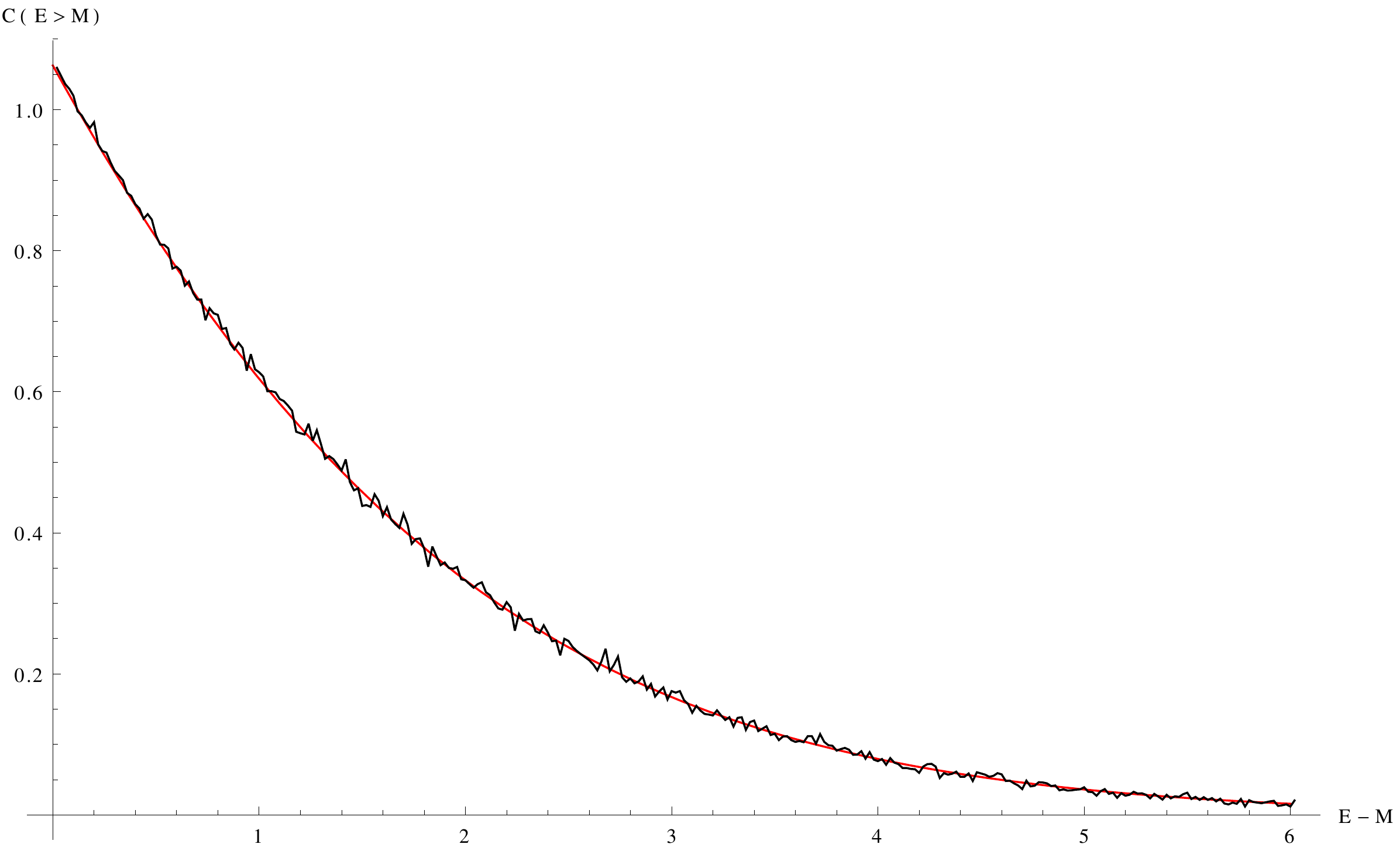}
\caption{Numerical estimate of $C(E>M)$ (black line), and its analytical 
approximation~\eqref{C(E)ThermApprox} (red line) for $N=10$ (energies are in Planck units).}
\label{PlotC}
\end{figure}

\setcounter{equation}{0}
We have checked the analytical approximation~\eqref{C(E)ThermApprox} by
computing directly some of the spectral coefficients~\eqref{Cseries} by means
of a Monte Carlo algorithm.
Fig.~\ref{SSim} shows the numerical estimates of $C_5=C_5(E-M)$ and
$C_{10}=C_{10}(E-M)$ along with their analytical approximation for $N=10$.
This comparison immediately shows that the coefficients $C_n$ indeed have
a ``Planckian'' shape, in agreement with their analytical approximation.
Finally, Fig.~\ref{PlotC} shows the whole coefficient $C(E>M)$ for $N=10$.
The same quantities for larger values of $N$, up to $N=100$ have also been computed
and the results are qualitatively the same.
\section{Useful integrals}
\setcounter{equation}{0}
\label{Integrals}
In this work we made use of the following results:
\be
\int_0^\infty \left(\frac{x}{e^x-1}\right)^2 \, \d x
\!\!&=&\!\!
\frac{1}{3}\left[\pi^2-6\,\zeta(3)\right]
\ ,
\\
\int_0^\infty \left(\frac{x}{e^x-1}\right)^2 \, x \, \d x
\!\!&=&\!\!
6\,\zeta(3)-\frac{\pi^4}{15}
\ ,
\\
\int_0^\infty \left(\frac{x}{e^x-1}\right)^2 \, x^2 \, \d x
\!\!&=&\!\!
\frac{4}{15}\left[\pi^4-90\,\zeta(5)\right]
\ .
\ee
%Since $\zeta(2)=\pi^2/6$ and $\zeta(4)=\pi^4/90$, we suggest
%the general rule to be
%\be
%\int_0^\infty \left(\frac{x}{e^x-1}\right)^2 \, x^n \, \d x
%\!\!&=&\!\!
%6[\zeta(n+2)-\zeta(n+3)]
%\ ,
%\ee
%for $n\geq0$.
%

%

\begin{thebibliography}{99}
%
%
\bibitem{DvaliGomez} 
G.~Dvali and C.~Gomez,
%``Quantum Compositeness of Gravity: Black Holes, AdS and Inflation'',
JCAP {\bf 01}, 023 (2014);
% arXiv:1312.4795 [hep-th];
``Black Hole's Information Group'',
arXiv:1307.7630;
%``Black Holes as Critical Point of Quantum Phase Transition,''
Eur.\ Phys.\ J.\ C {\bf 74}, 2752 (2014);
%[arXiv:1207.4059 [hep-th]];
%``Black Hole's 1/N Hair,''
Phys.\ Lett.\ B {\bf 719}, 419 (2013);
%[arXiv:1203.6575 [hep-th]];
%``Landau-Ginzburg Limit of Black Hole's Quantum Portrait: Self Similarity and Critical Exponent,''
Phys.\ Lett.\ B {\bf 716}, 240 (2012);
%[arXiv:1203.3372 [hep-th]];
%``Black Hole's Quantum N-Portrait,''
Fortsch.\ Phys.\  {\bf 61}, 742 (2013);
%[arXiv:1112.3359 [hep-th]];
G.~Dvali, C.~Gomez and S.~Mukhanov,
``Black Hole Masses are Quantized,''
arXiv:1106.5894 [hep-ph].
%
\bibitem{Bekenstein:1997bt}
J.~D.~Bekenstein,
``Quantum black holes as atoms,''
gr-qc/9710076.
%%CITATION = GR-QC/9710076;
%120 citations counted in INSPIRE as of 08 Jan 2015
%\cite{Casadio:2013hja}
%
\bibitem{Flassig:2012re}
D.~Flassig, A.~Pritzel and N.~Wintergerst,
%``Black Holes and Quantumness on Macroscopic Scales,''
Phys.~Rev.~D {\bf 87} (2013) 084007.
%[arXiv:1212.3344].
%%CITATION = ARXIV:1212.3344;
%9 citations counted in INSPIRE as of 21 Feb 2014
%
\bibitem{Casadio:2013hja}
R.~Casadio and A.~Orlandi,
%``Quantum Harmonic Black Holes,''
JHEP {\bf 1308} (2013) 025.
%[arXiv:1302.7138 [hep-th]].
%%CITATION = ARXIV:1302.7138;
%8 citations counted in INSPIRE as of 21 Feb 2014
%
\bibitem{Kuhnel:2014oja}
F.~Kuhnel,
%``Bose-Einstein Condensates with Derivative and Long-Range Interactions as Set-Ups for Analog Black Holes,''
Phys.~Rev.~D {\bf 90}, no. 8, 084024 (2014);
%[arXiv:1312.2977 [gr-qc]]
%%CITATION = ARXIV:1312.2977;
%4 citations counted in INSPIRE as of 23 Jan 2015
F.~Kuhnel and B.~Sundborg,
``Modified Bose-Einstein Condensate Black Holes in d Dimensions,''
arXiv:1401.6067 [hep-th];
%%CITATION = ARXIV:1401.6067;
%2 citations counted in INSPIRE as of 23 Jan 2015
F.~Kuhnel and B.~Sundborg,
%``High-Energy Gravitational Scattering and Bose-Einstein Condensates of Gravitons,''
JHEP {\bf 1412}, 016 (2014);
%[arXiv:1406.4147 [hep-th]];
%%CITATION = ARXIV:1406.4147;
%4 citations counted in INSPIRE as of 23 Jan 2015
F.~Kuhnel and B.~Sundborg,
%``Decay of Graviton Condensates and their Generalizations in Arbitrary Dimensions,''
Phys.~Rev.~D {\bf 90}, no. 6, 064025 (2014).
%[arXiv:1405.2083 [hep-th]];
%%CITATION = ARXIV:1405.2083;
%4 citations counted in INSPIRE as of 23 Jan 2015
%
\bibitem{mueckPT}
W.~M{\"u}ck and G.~Pozzo,
%``Quantum Portrait of a Black Hole with P\"oschl-Teller Potential,''
JHEP {\bf 1405}, 128 (2014).
% [arXiv:1403.1422 [hep-th]].
%%CITATION = ARXIV:1403.1422;%%
%1 citations counted in INSPIRE as of 04 Feb 2015
%
%\cite{Hofmann:2014jya}
\bibitem{Hofmann:2014jya}
S.~Hofmann and T.~Rug,
``A Quantum Bound-State Description of Black Holes,''
arXiv:1403.3224 [hep-th].
%%CITATION = ARXIV:1403.3224;
%5 citations counted in INSPIRE as of 15 Dec 2014
%
\bibitem{Casadio:2014vja}
R.~Casadio, A.~Giugno, O.~Micu and A.~Orlandi,
%``Black holes as self-sustained quantum states, and Hawking radiation,''
Phys.\ Rev.\ D {\bf 90}, 084040 (2014).
%[arXiv:1405.4192 [hep-th]].
%%CITATION = ARXIV:1405.4192;%%
%5 citations counted in INSPIRE as of 31 Mar 2015
%
\bibitem{winter} 
V.~F.~Foit and N.~Wintergerst,
``Self-similar Evaporation and Collapse in the Quantum Portrait of Black Holes,''
arXiv:1504.04384 [hep-th].
%%CITATION = ARXIV:1504.04384;%%
%
%\cite{Binetruy:2012kx}
\bibitem{Binetruy:2012kx}
P.~Binetruy,
``Vacuum energy, holography and a quantum portrait of the visible Universe,''
arXiv:1208.4645 [gr-qc].
%%CITATION = ARXIV:1208.4645;
%5 citations counted in INSPIRE as of 21 Feb 2014
%
%\cite{Kuhnel:2014gja}
\bibitem{Kuhnel:2014gja} 
F.~Kuhnel,
``Thoughts on the Vacuum Energy in the Quantum N-Portrait,''
arXiv:1408.5897 [gr-qc].
%%CITATION = ARXIV:1408.5897;%%
%
\bibitem{Dvali:2013eja} 
G.~Dvali and C.~Gomez,
%``Quantum Compositeness of Gravity: Black Holes, AdS and Inflation,''
JCAP {\bf 1401}, no. 01, 023 (2014);
%[arXiv:1312.4795 [hep-th]];
%%CITATION = ARXIV:1312.4795;%%
%22 citations counted in INSPIRE as of 31 Mar 2015
G.~Dvali and C.~Gomez,
``Quantum Exclusion of Positive Cosmological Constant?,''
arXiv:1412.8077 [hep-th].
%%CITATION = ARXIV:1412.8077;%%
%2 citations counted in INSPIRE as of 31 Mar 2015 
 %
\bibitem{Casadio:2015xva} 
R.~Casadio, F.~Kuhnel and A.~Orlandi,
``Consistent Cosmic Microwave Background Spectra from Quantum Depletion,''
arXiv:1502.04703 [gr-qc];
%%CITATION = ARXIV:1502.04703;%%
F.~Kuhnel, M.~Sandstad,
``Corpuscular consideration of eternal inflation''
arXiv:1504.02377.
%
\bibitem{HWF}
R.~Casadio,
``Localised particles and fuzzy horizons: A tool for probing Quantum Black Holes,''
arXiv:1305.3195 [gr-qc];
%%CITATION = ARXIV:1305.3195;%%
%10 citations counted in INSPIRE as of 31 Mar 2015
R.~Casadio and F.~Scardigli,
%``Horizon wave-function for single localized particles: GUP and quantum black hole decay,''
Eur.\ Phys.\ J.\ C {\bf 74}, no. 1, 2685 (2014)
[arXiv:1306.5298 [gr-qc]];
%%CITATION = ARXIV:1306.5298;%%
%14 citations counted in INSPIRE as of 31 Mar 2015
R.~Casadio,
``Horizons and non-local time evolution of quantum mechanical systems,''
arXiv:1411.5848 [gr-qc], to appear in EPJ~C.
%%CITATION = ARXIV:1411.5848;%%
%3 citations counted in INSPIRE as of 31 Mar 2015
%
\bibitem{davidson} 
A.~Davidson and B.~Yellin,
%``Quantum Black Hole Wave Packet: Average Area Entropy and Temperature Dependent Width,''
Phys. Lett. B, 267 (2014);
%[arXiv:1404.5729 [gr-qc]].
  %%CITATION = ARXIV:1404.5729;%%
  %1 citations counted in INSPIRE as of 04 Jun 2015
A.~Davidson and B.~Yellin,
``Thermal Hawking Broadening and Statistical Entropy of Black Hole Wave Packet,''
arXiv:1306.6403 [gr-qc].
%%CITATION = ARXIV:1306.6403;%%
%
\bibitem{stephani}
H.~Stephani,
``Relativity: An introduction to special and general relativity,''
Cambridge University Press, Cambridge, UK (2004).
%3 citations counted in INSPIRE as of 04 Jun 2015
%
\bibitem{micro}
B.~Harms and Y.~Leblanc,
%``Statistical mechanics of black holes,''
Phys.\ Rev.\ D {\bf 46}, 2334 (1992);
%  [hep-th/9205021].
%%CITATION = HEP-TH/9205021;%%
%62 citations counted in INSPIRE as of 14 Apr 2015
R.~Casadio and B.~Harms,
%``Microfield dynamics of black holes,''
Phys.\ Rev.\ D {\bf 58}, 044014 (1998);
%[gr-qc/9712017].
%%CITATION = GR-QC/9712017;%%
%47 citations counted in INSPIRE as of 14 Apr 2015
%``Microcanonical description of (micro) black holes,''
Entropy {\bf 13}, 502 (2011).
%  [arXiv:1101.1384 [hep-th]].
%%CITATION = ARXIV:1101.1384;%%
%12 citations counted in INSPIRE as of 14 Apr 2015
%
\bibitem{RN}
R.~Casadio, O.~Micu and D.~Stojkovic,
``Horizon Wave-Function and the Quantum Cosmic Censorship,''
arXiv:1503.02858 [gr-qc];
%%CITATION = ARXIV:1503.02858;%%
%1 citations counted in INSPIRE as of 15 Apr 2015
%R.~Casadio, O.~Micu and D.~Stojkovic,
``Inner Horizon of the Quantum Reissner-Nordstr\"om Black Holes,''
arXiv:1503.01888 [gr-qc].
%%CITATION = ARXIV:1503.01888;%%
%1 citations counted in INSPIRE as of 15 Apr 2015
%
\end{thebibliography}
\end{document}